\documentclass{webofc}
\usepackage[varg]{txfonts}   
\def\MS7{MS 0735.6+7421}
\begin{document}
\title{Sunyaev Zel'dovich effect in galaxy clusters cavities: thermal or non-thermal origin?}
%
%

\author{\firstname{Paolo} \lastname{Marchegiani}\inst{1,2}\fnsep\thanks{\email{paolo.marchegiani@uniroma1.it}}
}

\institute{Dipartimento di Fisica, Universit\`a Sapienza, P.le Aldo Moro 2, 00185, Roma, Italy 
\and
           Wits Centre for Astrophysics, School of Physics, University of the Witwatersrand, 1 Jan Smuts Avenue, Johannesburg 2050, South Africa 
          }

\abstract{%
Several galaxy clusters host X-ray cavities, often filled with relativistic electrons emitting in the radio band.
In the cluster MS 0735.6+7421 the cavities have been detected through the Sunyaev Zel'dovich (SZ) effect, but it has not been possible to determine if this effect is thermal (produced by a very high temperature gas filling the cavity) or non-thermal (produced by the relativistic electrons that produce the diffuse radio emission detected in the cavity).
In this paper we discuss the role of the density of the high temperature gas inside the cavities in determining whether the dominant SZ effect is the thermal or the non-thermal one, and how it can be possible to distinguish between the two possibilities, discussing the role of observations at higher energy bands.
}
\maketitle
\section{Introduction}
\label{intro}

Several relaxed galaxy clusters show the presence inside their Intra Cluster Medium (ICM) of couples of cavities in the X-ray emission, usually located in opposite directions compared to the central Brightest Cluster Galaxy \cite{Fabian2012,Gitti2012}. The origin of these cavities is probably due to the mechanical power produced by AGN jets that is converted to heat during the expansion of the lobe in the ICM \cite{McNamara2007}. Many X-ray cavities host a diffuse radio emission \cite{Birzan2020}, showing that relativistic electrons and magnetic fields are present in them. However, it has been found that the energy stored in these relativistic electrons, determined under the minimum energy assumption, is well below the energy necessary to inflate the cavity \cite{Ito2008}, indicating that another component inside the cavity should be dominant in energy. This dominant component can be a high temperature thermal gas with a temperature of the order of $kT\sim100$ keV or higher according to hydrodynamic simulation \cite{Sternberg2009,Prokhorov2012}, or a population of relativistic protons.

While radio observations are suitable to constrain the high energy part of the relativistic electrons spectrum and the magnetic field, they are not able to provide information on the low energy part of the electrons spectrum, and on the thermal gas inside the cavity. The Sunyaev-Zel'dovich effect (SZE) produced inside the cavity, i.e. the distortion of the Cosmic Microwave Background (CMB) spectrum by inverse Compton scattering (ICS) \cite{Sunyaev1972}, has instead been proposed as a probe to constrain the low energy electrons or the thermal gas properties \cite{Colafrancesco2005,Pfrommer2005,Prokhorov2010}. With this kind of information it is possible to constrain the energetic content of the different components inside the cavity, because the low energy part of the electrons spectrum is the dominant one in determining the total energy stored in them \cite{Colafrancesco2011}.

The galaxy cluster \MS7 hosts a couple of X-ray cavities \cite{McNamara2005,Gitti2007,Vantyghem2014}, where a diffuse radio emission filling the cavities is also observed \cite{Birzan2008,Birzan2020}. These cavities have also been observed through the SZE at 30 GHz using the Combined Array for Research in Millimeter-wave Astronomy (CARMA) interferometer \cite{Abdulla2019}, which detected a clear deficit in the SZ signal in the direction of the cavities. From these observations, it has not been possible to determine if the dominant component inside the cavities is thermal or non-thermal, because the data were compatible with both a non-thermal electrons population having a low value of the minimum moment ($p_1\sim 1 - 10$), and with a thermal population with very high temperature of $kT\sim1000$ keV, or even higher (see fig.8 in \cite{Abdulla2019}).

In a recent paper \cite{Marchegiani2021} we discussed this problem, pointing up that these two different populations, i.e. the non-thermal electrons and the high temperature gas, are linked each other because of the effect of the Coulomb interactions. In particular, a higher density of the thermal gas implies a higher rate of the Coulomb losses, which have the effect of reducing the density of the low energy non-thermal electrons and therefore of flattening their spectrum, and as a consequence of reducing the intensity of the non-thermal SZE. Therefore, this effect should be taken into account when interpreting the results of observations of the SZE inside the cluster cavities in terms of thermal or non-thermal populations.

In this paper, we summarize the results obtained in \cite{Marchegiani2021}, and expand further the discussion about how it can be possible to discriminate between the thermal and the non-thermal origin of the SZE inside the cavities, also considering the expected SZE in spectral bands with higher energies than the microwave one.

\section{Modeling the spectrum of non-thermal electrons}
\label{sec-1}

The time evolution of the non-thermal electrons spectrum $N_e(p,t)$, written as a function of the time $t$ and the normalized momentum $p=\beta\gamma$, is given the following equation \cite{Schlickeiser2002}:
\begin{equation}
\frac{\partial N_e(p,t)}{\partial t}  =  \frac{\partial}{\partial p} \left[ \left(-\frac{2}{p}D_{pp}+\sum_i b_i (p) \right) N_e(p,t) + D_{pp}\frac{\partial N_e(p,t)}{\partial p}\right],
\label{evol.spettro}
\end{equation}
where $D_{pp}$ is the diffusion term in the momentum space, associated to Fermi-II acceleration processes, which we model as $D_{pp}=\chi p^2 /4$ so that the characteristic acceleration time is $\tau_{acc}=p^2/4D_{pp}=\chi^{-1}$ \cite{Brunetti2007}. The energy loss term $b_i(p)$ is given by the sum of the Coulomb losses, $b_{Coul}\propto n_{th}$ \cite{Gould1972}, where $n_{th}$ is the density of the thermal gas, of the radiative losses (synchrotron and inverse Compton scattering), $b_{rad}\propto (B^2+B^2_{ICS}) p^2$ \cite{Winner2019}, and of the adiabatic expansion losses, $b_{ad} \propto (dV/dt) p$, where we assume that the volume is changing with a Sedov-like expansion $V(t)\propto (t/t_0)^{6/5}$ \cite{Ensslin2001}.

We assume an initial electrons spectrum with a power law shape, $N_e(p,t_0)=k_0 p^{-s}$ with $s=2.7$, and initial values of the magnetic field and the thermal density so that at the present time they are equal respectively to the equipartion value $B=4.7$ $\mu$G \cite{Birzan2008} and to a reference value chosen in the range $10^{-6}-10^{-3}$ cm$^{-3}$. We therefore calculate the electrons spectrum after a time of the order of the age of the bubble, estimated to be $\sim160$ Myr from X-ray observations \cite{Vantyghem2014} and MHD simulations \cite{Ehlert2019}, and choose the electrons normalization $k_0$ and the Fermi-II acceleration parameter $\chi$ in order to match the observed spectrum of the radio emission, with the best parameters values resulting to be $k_0=1.5\times10^{-3}$ cm$^{-3}$ and $\chi=3.5\times10^{-16}$ s$^{-1}$ \cite{Marchegiani2021}. 

\begin{figure}[t]
\begin{center}
\includegraphics[scale=0.35]{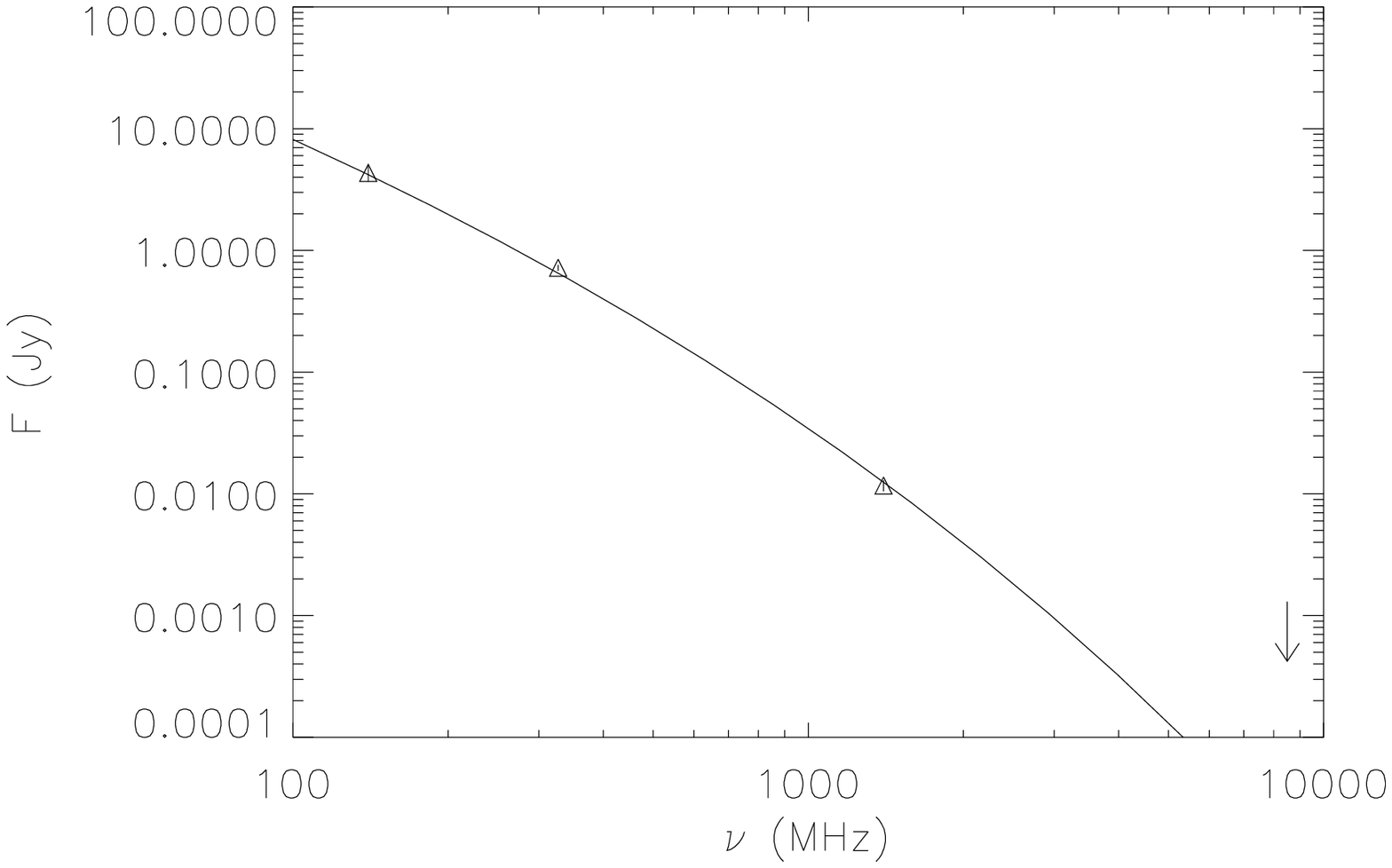}
\includegraphics[scale=0.35]{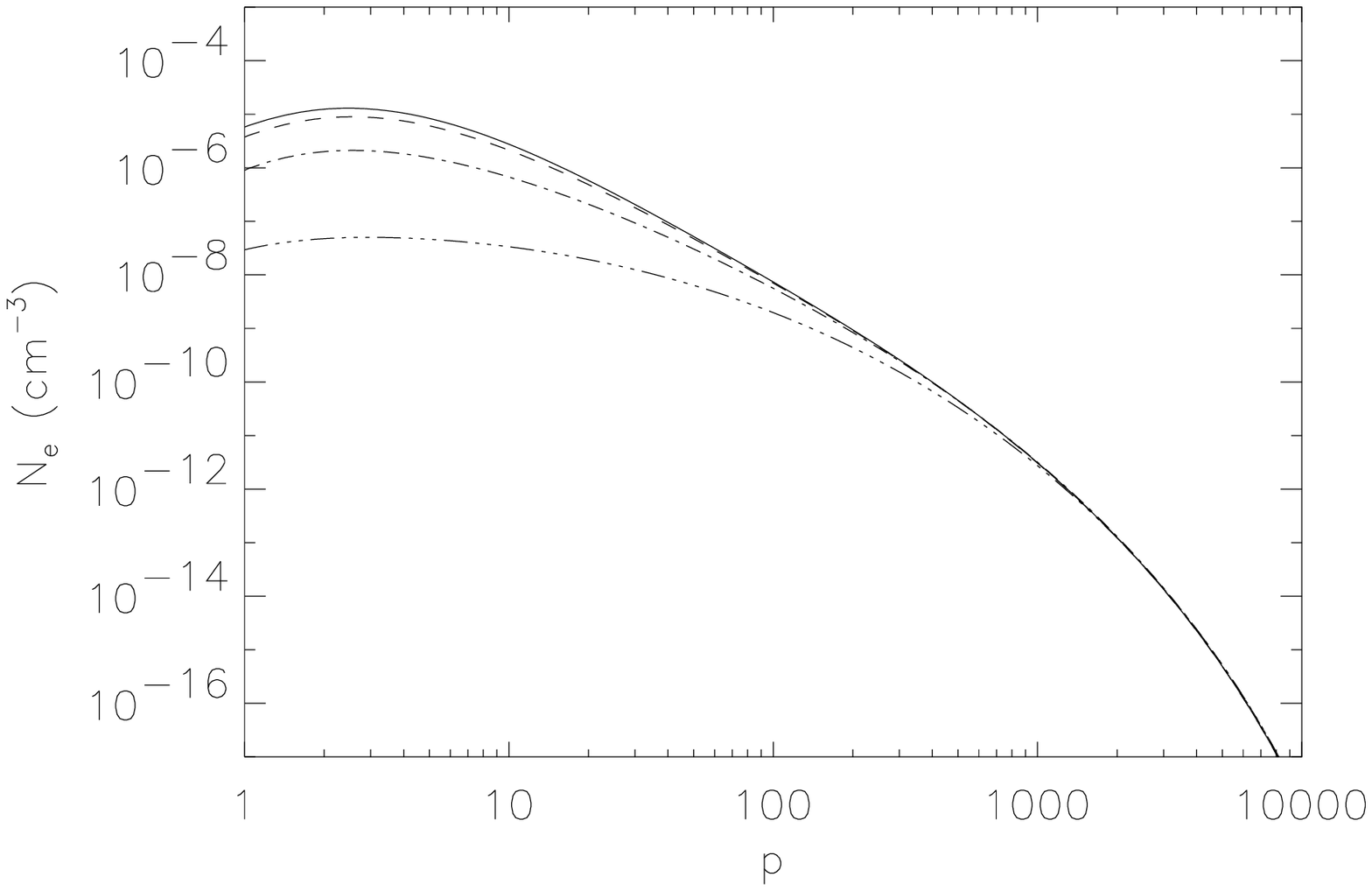}
\caption{Radio spectrum after 160 Myr from the injection  (left panel) and the corresponding non-thermal electrons spectra (right panel) for four values of the thermal density: $10^{-6}$ (solid line), $10^{-5}$ (dashed line), $10^{-4}$ (dot-dashed line), and $10^{-3}$ (three dots-dashed line) cm$^{-3}$. Figures are from \cite{Marchegiani2021}.}
\label{fig-radio}
\end{center}
\end{figure}

In Fig.\ref{fig-radio}, left panel, the resulting radio spectrum, compared with the observed data \cite{Birzan2008,Birzan2020}, is shown for four values of the thermal density inside the cavity, while the electrons spectra for the same values of the thermal density are shown in the right panel. It is interesting to see that, while the value of the thermal density does not impact too much on the shape of the radio spectrum (in fact the relative differences between different models are smaller than 10\% in the whole frequency range between 100 MHz and 10 GHz), it heavily affects the spectrum of the electrons at low energies. As shown in the next Section, this fact has strong consequences on the SZE produced by these electrons.

\section{The non-thermal Sunyaev-Zel'dovich effect}

Using the electrons spectra calculated in the previous Section, it is possible to calculate the non-thermal SZE, using the full relativistic formalism \cite{Wright1979,Colafrancesco2003}, where the intensity of the SZE is given by
$\Delta I (x) =  \tau [J_1(x)-I_0(x)]$.
Here $x=h\nu/(k_B T_{0})$ is the frequency normalized at the CMB temperature $T_0$, $I_0(x)$ is the unperturbed CMB spectrum, $\tau$ is the electrons optical depth, proportional to the electrons density integrated along the line of sight, and $J_1(x)$ is calculated by convolving the CMB spectrum with the single scattering redistribution function, which is given by the convolution of the electrons spectrum with the Compton scattering redistribution function for an electron with a given energy \cite{Ensslin2000}.

\begin{figure}[t]
\begin{center}
\includegraphics[scale=0.35]{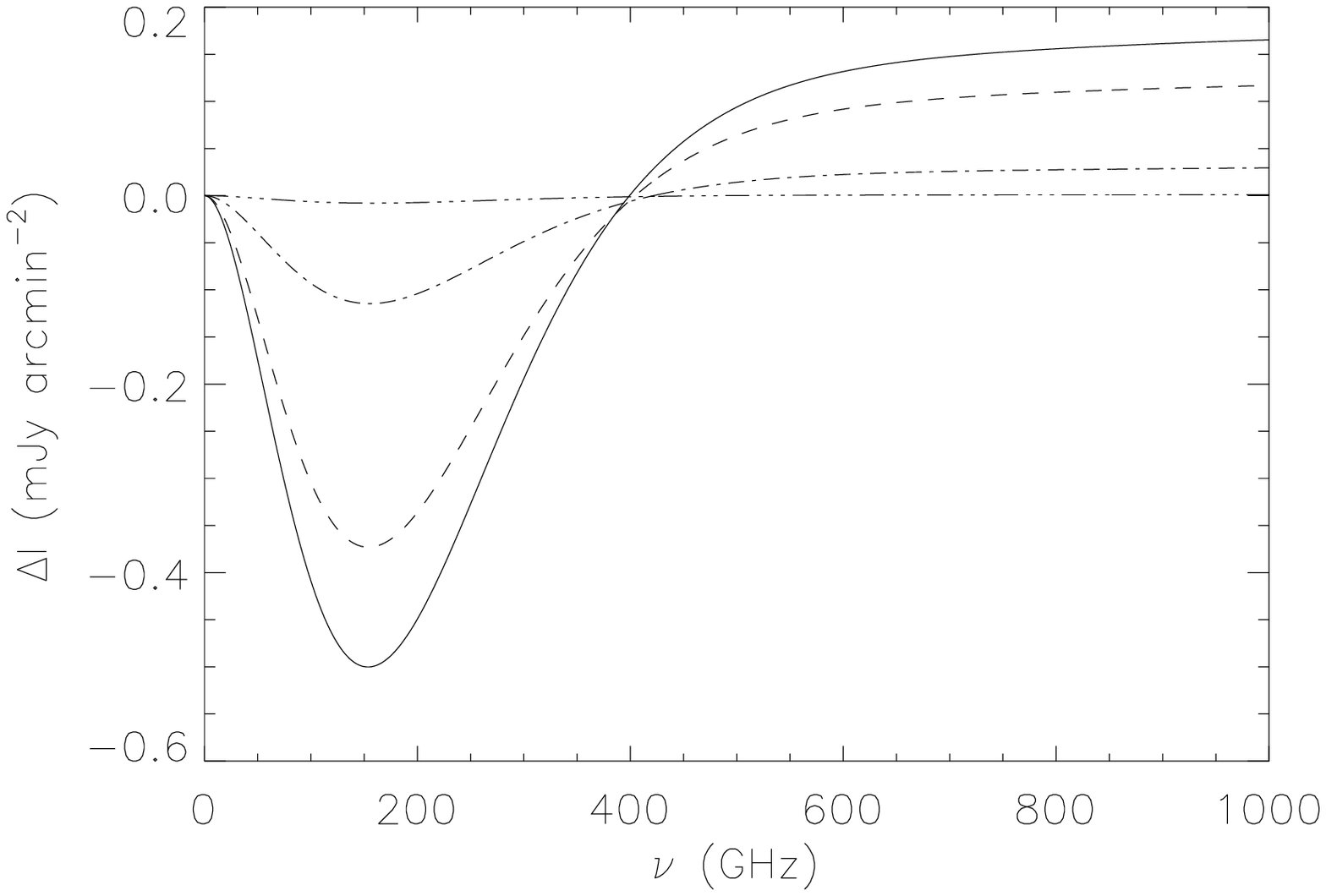}
\includegraphics[scale=0.35]{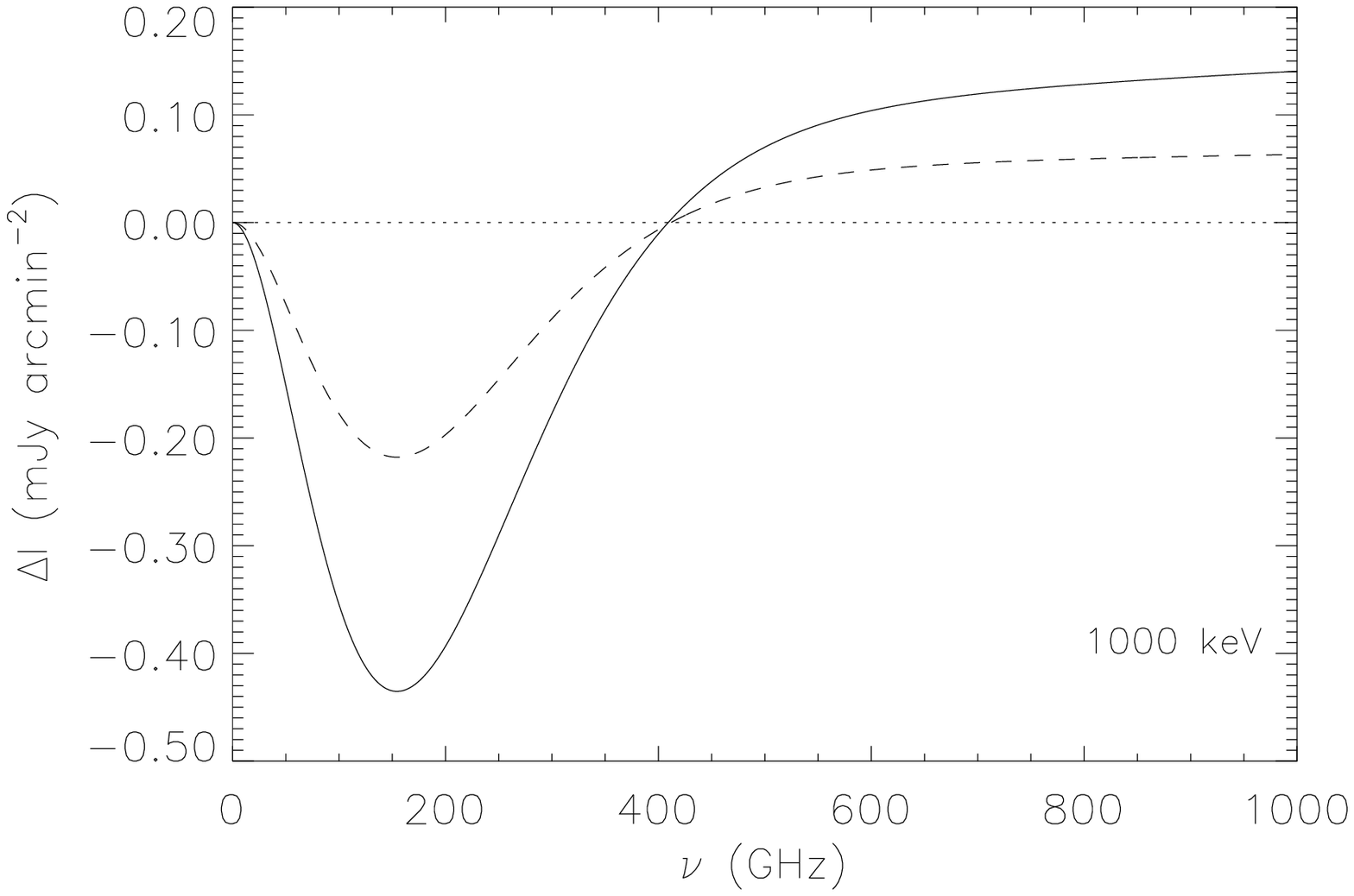}
\includegraphics[scale=0.35]{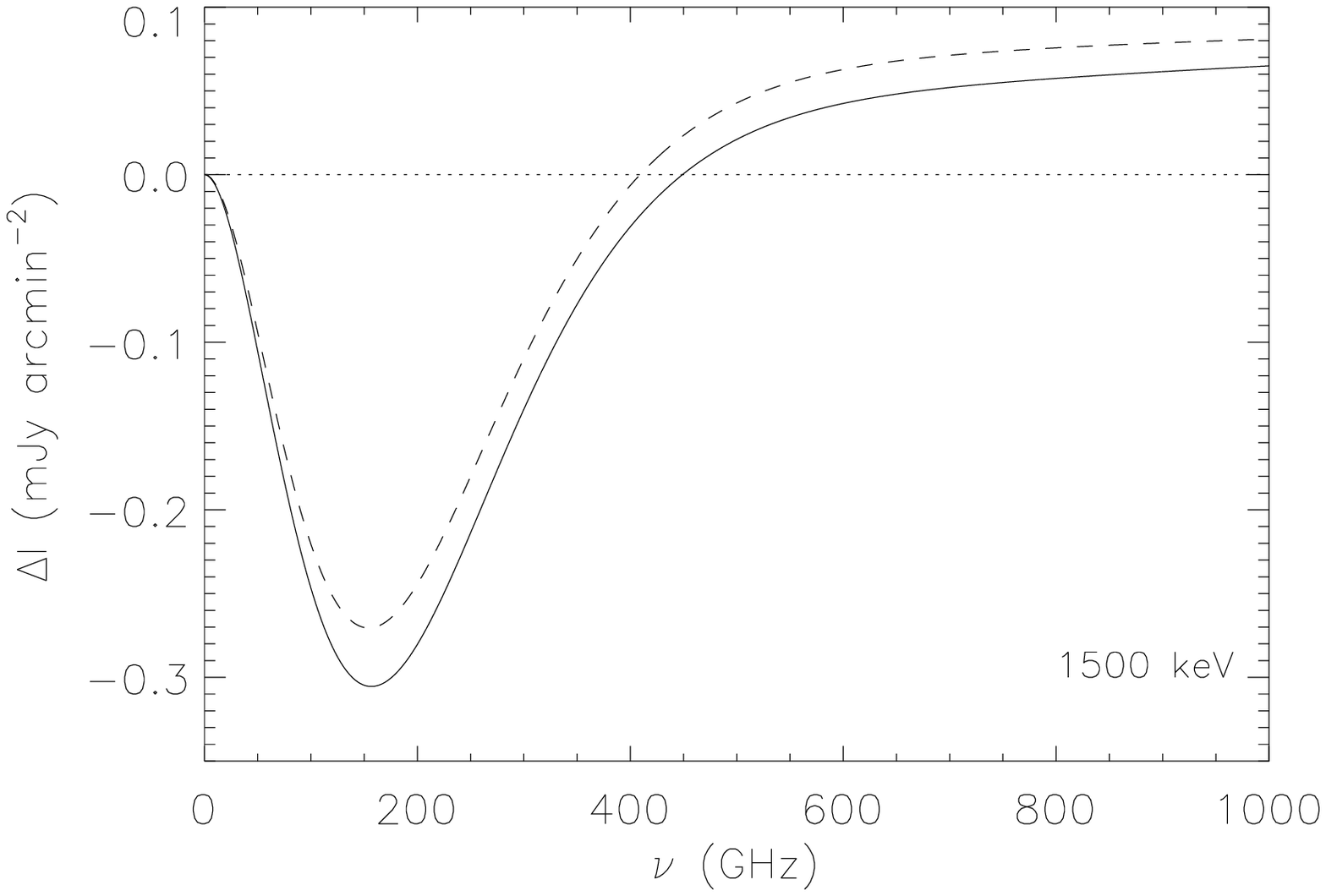}
\includegraphics[scale=0.35]{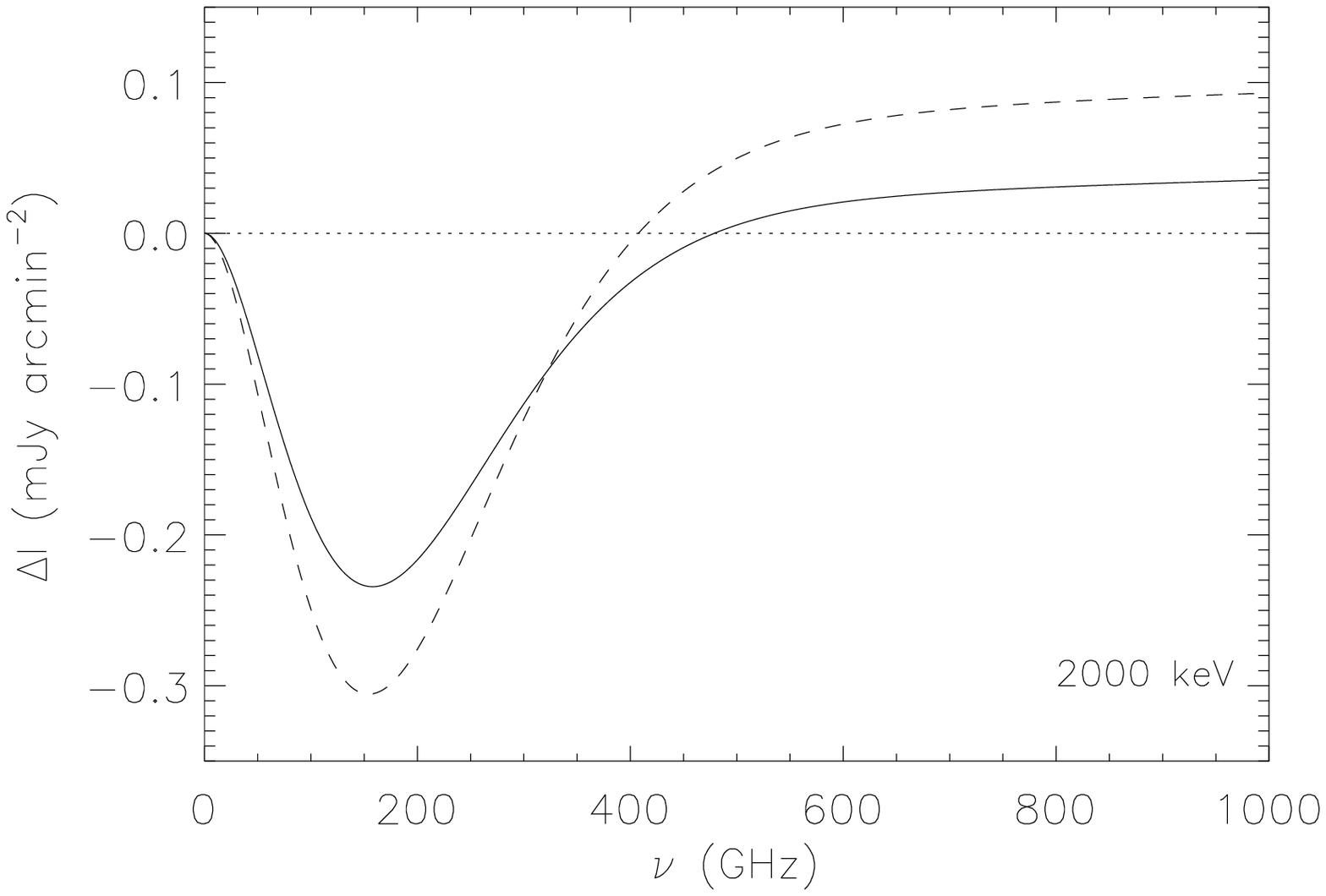}
\caption{First panel: non-thermal SZE produced by the electrons with the spectra shown in Fig.\ref{fig-radio}, with the same meaning of the line styles. Other panels: thermal SZE inside the cavities (solid lines) compared with the non-thermal SZE calculated for the corresponding values of the thermal density (dashed lines) for values of the temperature of 1000, 1500, and 2000 keV. The dotted line in last three panels indicates the zero level. Figures are from \cite{Marchegiani2021}.}
\label{fig-sz}
\end{center}
\end{figure}

The resulting SZE spectra for the four values of the thermal density that were used in the previous Section are shown in the first panel of Fig.\ref{fig-sz}. As it is possible to see, smaller values of the thermal density imply reduced Coulomb losses, and therefore a higher density of the non-thermal electrons at low energies and as a consequence a more intense SZE.

The SZE produced by the non-thermal electrons inside the cavity can be compared with the thermal SZE produced by the high temperature gas inside the same cavity. Since we don't have observational information about the properties of this gas, we can assume several values for its temperature, and assume for its density the values for which its pressure is equal to the one of the external ICM at the location of the cavity, $P_{cav}=6\times10^{-11}$ erg cm$^{-3}$ \cite{Gitti2007}, i.e. $n_{th}=P_{cav}/(kT)$. Using the full relativistic formalism, we calculate the thermal SZE for these temperatures and the optical depths derived from these values of the density, and the non-thermal SZE for the Coulomb losses rates provided by the same values of the thermal density. In the last three panels of Fig.\ref{fig-sz} the results for three values of the temperature are shown.

As it is possible to see, values of the temperature until $\sim1500$ keV allow to have quite high values of the thermal density, and this fact increases the intensity of the thermal SZE and at the same time decreases the one of the non-thermal SZE because of the Coulomb losses. For higher temperatures, we can expect that the non-thermal SZE should be dominant, while for lower temperatures the thermal effect can be dominant, even if it is possible that the thermal density is lower than the value assumed here, so it is not possible to state whether in this case the dominant effect is definitely the thermal one.

\section{Thermal or non-thermal SZE?}

In this Section we discuss how it can be possible to determine if the dominant SZE inside the cavity is the thermal or the non-thermal one, adding some other considerations as compared to the previous paper \cite{Marchegiani2021}, to which we refer instead for other details on the models, on the properties of the electrons populations used in the calculations, on the comparison between the SZE inside the cavity and the one of the surrounding ICM, and on the observational challenges to be faced in order to derive this kind of information.

As discussed in the previous paper, the angular size of the cavities in \MS7 is of the order of 1 arcmin; therefore instruments with angular resolution of the order of 10 arcsec like NIKA2 are well suited to resolve the SZ signal inside the cavities, and to distinguish it from the one coming from the rest of cluster. While NIKA2 can provide information at the frequencies of 150 and 260 GHz, a wider spectral coverage could be useful to better distinguish between a thermal and a non-thermal origin of the SZE; this can be done with instruments operating at 90 GHz, like Mustang-2 at the Green Bank Telescope (GBT) and Mistral at the Sardinia Radio Telescope (SRT), and at higher frequencies, like the planned space telescope Millimetron \cite{Kardashev2014}, with its planned angular resolution of a few arcsec and a spectral coverage until a few THz.

However, also a full spectral frequency coverage of the SZE inside the cavity might not be sufficient to determine if its origin is thermal or non-thermal. In fact, in Fig.\ref{fig-sz} we can see that for temperatures in the range 1000 -- 1500 keV the thermal and the non-thermal SZE are expected to have very similar spectral shapes (note for example that the crossover frequency in the top right panel of Fig.\ref{fig-sz} is basically the same for the two effects).

\begin{figure}[t]
\begin{center}
\includegraphics[scale=0.35]{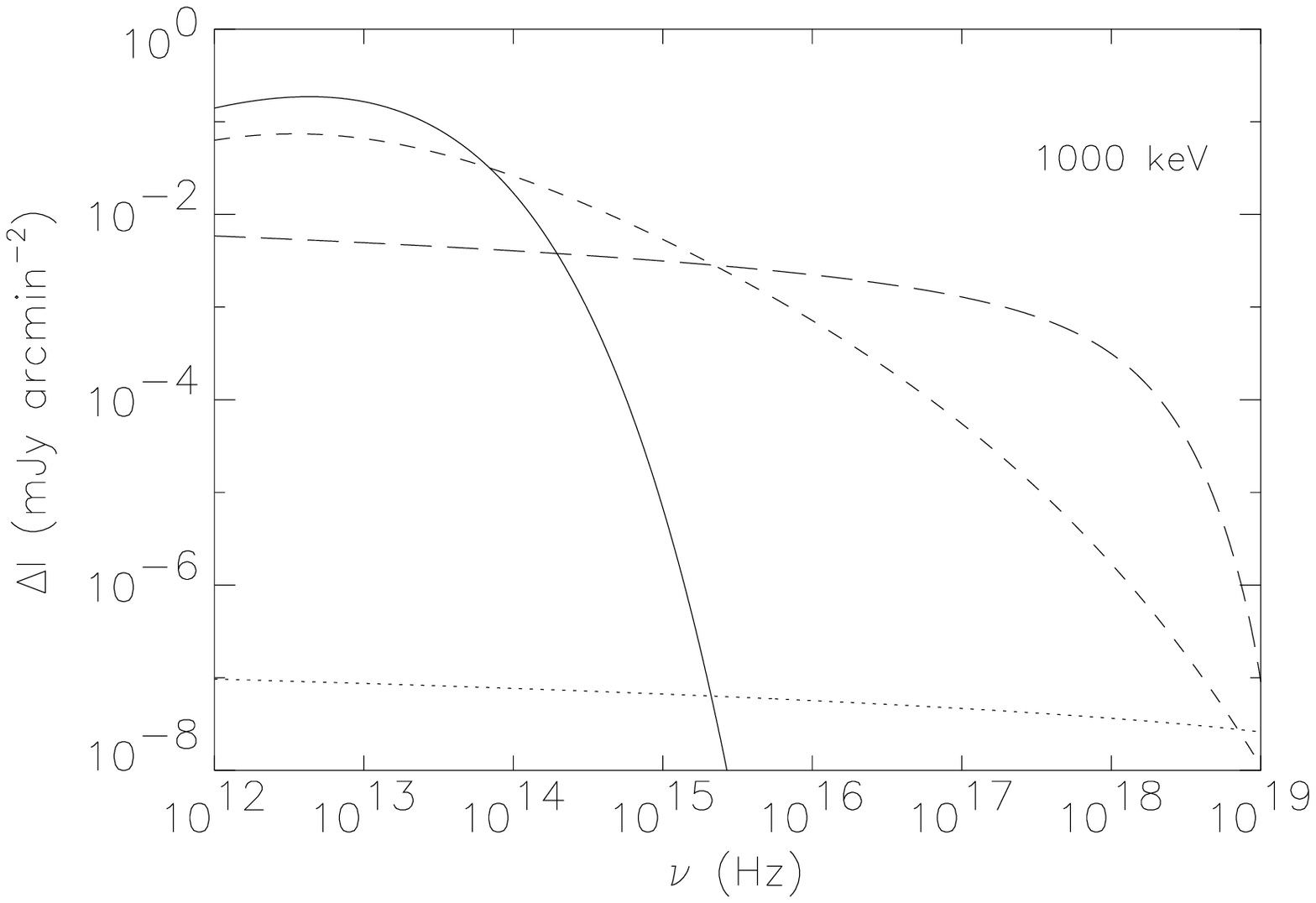}
\includegraphics[scale=0.35]{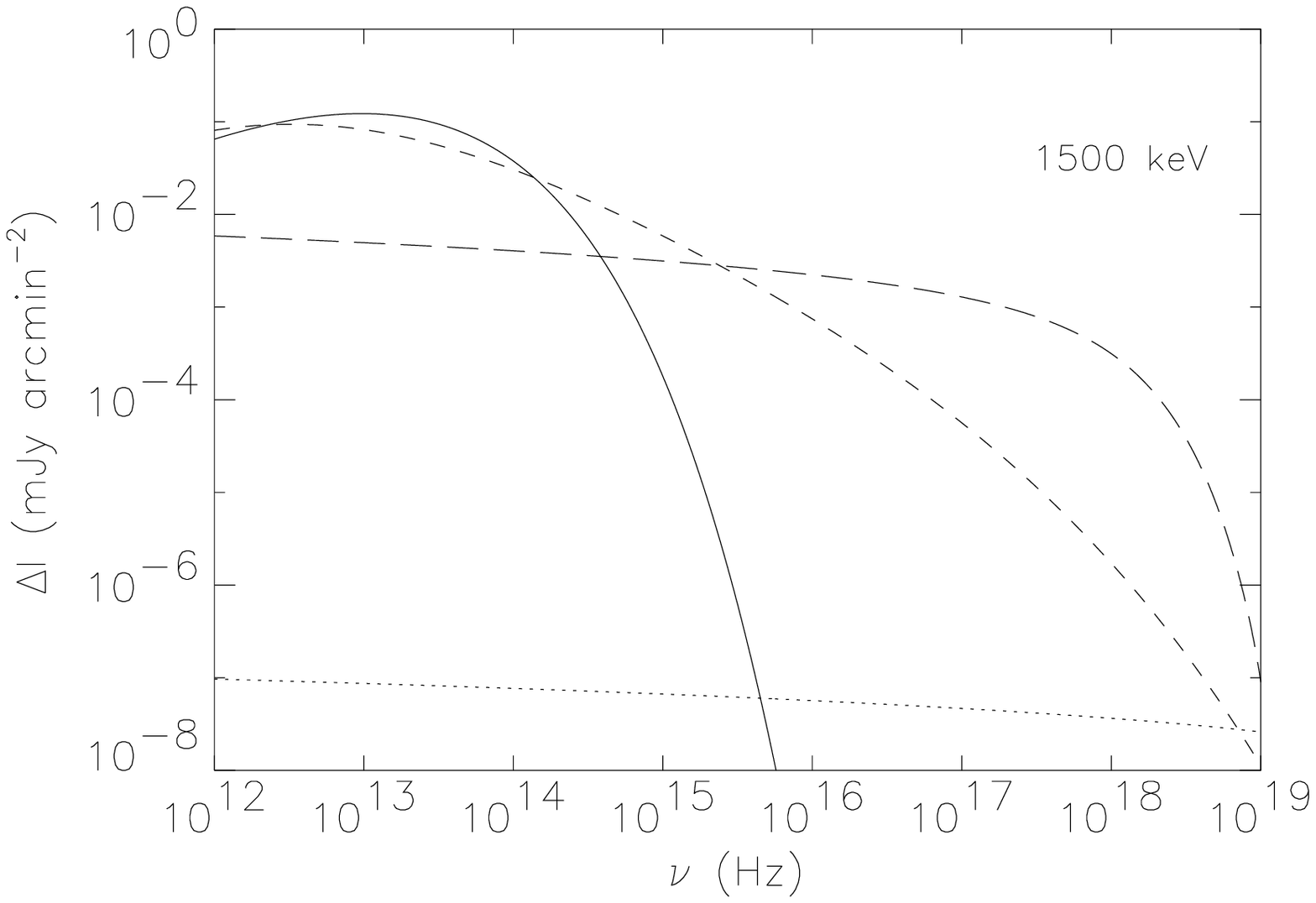}
\caption{SZE spectra at high frequencies for the thermal gas inside the cavity (solid lines) with temperatures of 1000 (left panel) and 1500 keV (right panel), compared with the SZE spectra of the non-thermal electrons (dashed lines), and with the bremsstrahlung of the high temperature gas inside the cavity (dotted line) and of the external ICM along the line of sight of the cavity center (long dashed lines).}
\label{fig-highe}
\end{center}
\end{figure}

A possible solution might be to look at the SZE emission at frequencies higher than the THz, where it should reach its maximum, depending on the properties of the electrons population. For example, it has been suggested that the Hard X-ray band can be good to distinguish between the ICS emission of the non-thermal electrons and the bremsstrahlung of the high temperature thermal gas, because the efficiency of the ICS process is much higher than the bremsstrahlung one \cite{Prokhorov2012}. In order to check this possibility, in Fig.\ref{fig-highe} we compare the thermal and non-thermal SZE spectra calculated at high frequencies, from the THz until $10^{19}$ Hz (corresponding to a photon energy of $\sim40$ keV), for values of the gas temperature inside the cavity of 1000 and 1500 keV.

As it is possible to see, the spectra of the thermal ad the non-thermal SZE are quite similar until frequencies of $\nu\sim10^{14}$ Hz, while they start to diverge at higher frequencies, with the thermal SZE falling down quite abruptly, and the non-thermal SZE continuing instead with a power-law shape until the Hard X-ray band, with a progressive steepening due to the corresponding high energy steepening of the electrons spectrum, visible in the right panel of Fig.\ref{fig-radio}. Therefore, a hypothetical detection of the SZE inside the cavity at frequencies $>10^{14}$ Hz, i.e. from the near infrared band to higher frequencies, should be attributed to the non-thermal electrons rather than to the high temperature gas. Such an effect should appear as a positive diffuse emission spatially coincident with the cavity.

In Fig.\ref{fig-highe} we also show an estimate of the surface brightness in the direction of the cavity center due to the thermal bremsstrahlung of the high temperature gas inside the cavity and of the external ICM present along the line of sight. This last quantity has been calculated assuming the presence of a gas with constant temperature of 5.5 keV and with the double beta radial profile derived from X-ray observations (see \cite{Marchegiani2021} for details), extended until the virial radius of 2.23 Mpc \cite{Gitti2007}, along the line of sight at the distance of 150 kpc from the cluster center, and excluding from the integration along the line of sight the cavity region, approximated as a sphere with radius of 100 kpc. The bremsstrahlung produced by the high temperature gas inside the cavity results to be much lower than the one from the ICM, as expected because of its lower density, higher temperature, and smaller size. The bremsstrahlung produced by the ICM is instead dominant on the SZE produced by the non-thermal electrons at frequencies higher than $\sim2\times10^{15}$ Hz, i.e. in the far ultraviolet band. 

Therefore there is a spectral region, going from the near infrared to the far ultraviolet, where the non-thermal SZE should be the dominant diffuse component inside the cavity. Should a diffuse emission spatially coincident with the cavity in these spectral bands be detected, it should therefore be attributed to the non-thermal SZE. In particular, the most favorable spectral region results to be at a wavelength around 400 nm, i.e. between the near ultraviolet and the visible spectral regions. Instruments that can be suitable to detect such an emission might be UVOT aboard the \textit{Swift} satellite, or the WFC3 camera aboard the Hubble Space Telescope; in order to establish the effective possibility to detect this signal, a dedicated study would be required.

\vspace{0.5cm}

\small{
\section*{Acknowledgements}

The Author thanks the organizers and the participants of the ``mm Universe @NIKA2'' Conference for very interesting and stimulating talks and discussions. The Author also thanks the Referee for useful comments that helped to improve the quality of the paper, and A. Maselli for his help and suggestions.}

%

\begin{thebibliography}{}
%

\bibitem{Fabian2012}
Fabian A.C., 2012, ARA\&A, 50, 455

\bibitem{Gitti2012}
Gitti M., Brighenti F., McNamara B.R. 2012, AdAst, 2012, 950641

\bibitem{McNamara2007}
McNamara B. R., Nulsen, P. E. J., 2007, ARA\&A, 45, 117

\bibitem{Birzan2020}
Birzan L. et al., 2020, MNRAS, 496, 2613

\bibitem{Ito2008}
Ito H., Kino M., Kawakatu N., et al., 2008, ApJ, 685, 828

\bibitem{Sternberg2009}
Sternberg A., Soker N., 2009, MNRAS, 398, 422 

\bibitem{Prokhorov2012}
Prokhorov D.A., Moraghan A., Antonuccio-Delogu V., Silk J., 2012, MNRAS, 425, 1753

\bibitem{Sunyaev1972}
Sunyaev R.A., Zel'dovich Ya.B., 1972, CoASP, 4, 173

\bibitem{Colafrancesco2005}
Colafrancesco S., 2005, A\&A, 435, L9

\bibitem{Pfrommer2005}
Pfrommer C., En\ss lin, T.A., Sarazin C.L., 2005, A\&A, 430, 799

\bibitem{Prokhorov2010}
Prokhorov D.A., Antonuccio-Delogu V., Silk J., 2010, A\&A, 520, A106

\bibitem{Colafrancesco2011}
Colafrancesco S.,  Marchegiani P., 2011, A\&A, 535, A108

\bibitem{McNamara2005}
McNamara B.R., Nulsen P.E.J., Wise M.W., et al., 2005, Nature, 433, 45

\bibitem{Gitti2007}
Gitti M., McNamara B.R., Nulsen P.E.J., Wise M.W., 2007, ApJ, 660, 1118

\bibitem{Vantyghem2014}
Vantyghem A.N., McNamara B.R., Russell H.R., et al., 2014, MNRAS, 442, 3192

\bibitem{Birzan2008}
Birzan L., McNamara B.R., Nulsen P.E.J., et al., 2008, ApJ, 686, 859

\bibitem{Abdulla2019}
Abdulla Z. et al., 2019, ApJ, 871, 195

\bibitem{Marchegiani2021}
Marchegiani P., 2021, MNRAS, 503, 4183

\bibitem{Schlickeiser2002}
Schlickeiser R., 2002, \textit{Cosmic Ray Astrophysics}, Springer-Verlag, Berlin

\bibitem{Brunetti2007}
Brunetti G., Lazarian A., 2007, MNRAS, 378, 245

\bibitem{Gould1972}
Gould R.J., 1972, Physica, 60, 145

\bibitem{Winner2019}
Winner G., Pfrommer C., Girichidis P., Pakmor R., 2019, MNRAS, 488, 2235

\bibitem{Ensslin2001}
En\ss lin T.A., Gopal-Krishna, 2001, A\&A, 366, 26

\bibitem{Ehlert2019}
Ehlert K., Pfrommer C., Weinberger R., Pakmor R., Springel V., 2019, ApJ, 872, L8

\bibitem{Wright1979}
Wright E.L., 1979, ApJ, 232, 348

\bibitem{Colafrancesco2003}
Colafrancesco S., Marchegiani P., Palladino E., 2003, A\&A, 397, 27

\bibitem{Ensslin2000}
En\ss lin T.A., Kaiser C.R., 2000, A\&A, 360, 417

\bibitem{Kardashev2014}
Kardashev N.S. et al., 2014, Physics-Uspekhi, 57, 12

\end{thebibliography}
%
%

\end{document}